\documentclass[9pt,conference]{IEEEtran}

\def\BibTeX{{\rm B\kern-.05em{\sc i\kern-.025em b}\kern-.08em
    T\kern-.1667em\lower.7ex\hbox{E}\kern-.125emX}}

\usepackage{graphicx}
\usepackage{amsmath,amssymb,amsfonts}   
\usepackage{glossaries}                 
\usepackage{enumitem}                   
\usepackage{makecell}                   
\usepackage{pifont}                     
\usepackage{cleveref}                   
\usepackage{color-edits}                
\usepackage{multirow}                   
\usepackage[table]{xcolor}              
\usepackage{float}                      
\usepackage{caption}                    
\usepackage{booktabs}                   
\usepackage{balance}                    
\usepackage{setspace}                   

\definecolor{lightgray}{HTML}{CCCCCC}

\newif\ifnb     
\nbfalse
\nbtrue

\newif\ifcom    
\comtrue

\newif\ifarx    
\arxfalse

\crefname{algocf}{Alg.}{Algs.}
\crefname{figure}{Fig.}{Figs.}
\crefname{equation}{Eq.}{Eqs.}

\addauthor[Patrick]{pi}{teal}
\renewcommand{\pi}[1]{\ifcom\picomment{#1}\fi}

\newcommand{\cmark}{\ding{52}}%
\newcommand{\xmark}{\ding{56}}%
\newcommand{\rone}{\raisebox{-0.125em}{\includegraphics[height=0.75em]{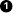}}}
\newcommand{\rtwo}{\raisebox{-0.125em}{\includegraphics[height=0.75em]{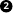}}}
\newcommand{\rthree}{\raisebox{-0.125em}{\includegraphics[height=0.75em]{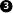}}}
\newcommand{\rfour}{\raisebox{-0.125em}{\includegraphics[height=0.75em]{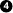}}}

\newcommand{\pkgaone}{\raisebox{-0.125em}{\includegraphics[height=0.75em]{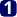}}}
\newcommand{\pkgatwo}{\raisebox{-0.125em}{\includegraphics[height=0.75em]{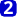}}}
\newcommand{\pkgbone}{\raisebox{-0.125em}{\includegraphics[height=0.75em]{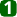}}}
\newcommand{\pkgbtwo}{\raisebox{-0.125em}{\includegraphics[height=0.75em]{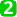}}}

\newcommand{\githuburl}{\ifnb https://github.com/spcl/PAISF \else https://github.com/anon8274/paisf (anonymized for blind review)\fi}

\newacronym{ci}{CI}{confidence interval}
\newacronym{llm}{LLM}{large language model}
\newacronym{moe}{MoE}{Mixture of Experts}
\newacronym{fft}{FFT}{fast Fourier transform}
\newacronym{phy}{PHY}{physical layer}
\newacronym{scb}{SCB}{simple cycle-breaking}
\newacronym{noc}{NoC}{network-on-chip}
\newacronym{ici}{ICI}{inter-chip interconnect}
\newacronym{pici}{ICI-P}{inter-chip interconnect with a passive substrate}
\newacronym{aici}{ICI-A}{inter-chip interconnect with an active interposer}
\newacronym{wsi}{WSI}{wafer-scale integration}
\newacronym{rf}{RF}{routing function}
\newacronym{sf}{SF}{selection function}
\newacronym{vc}{VC}{virtual channel}
\newacronym{ml}{ML}{machine learning}
\newacronym{io}{io}{in-order}
\newacronym{ooo}{ooo}{out-of-order}
\newacronym{rob}{ROB}{reorder buffer}
\newacronym{flit}{flit}{flow control unit}
\newacronym{cdg}{CDG}{channel dependency graph}
\newacronym{ilp}{ILP}{integer linear program}

\newacronym{m}{M}{2D mesh}
\newacronym{t}{T}{2D torus}
\newacronym{ft}{FT}{folded torus}
\newacronym{hc}{HC}{hypercube}
\newacronym{fbf}{FBF}{flattened butterfly}
\newacronym{hm}{HM}{HexaMesh}
\newacronym{fht}{FHT}{FoldedHexaTorus}
\newacronym{ks}{KS}{Kite Small}
\newacronym{km}{KM}{Kite Medium}
\newacronym{kl}{KL}{Kite Large}
\newacronym{dbf}{DBF}{double butterfly}
\newacronym{bd}{BD}{ButterDonut}

\newacronym{isf}{ISF}{in-order selection function}
\newacronym{paisf}{PAISF}{partially adaptive in-order selection function}
\newacronym{det}{DET}{deterministic selection function}
\newacronym{rnd}{RND}{random selection function}
\newacronym{adp}{ADP}{adaptive selection function}

\newacronym{lash}{LASH}{layered shortest path}

\begin{document}

\title{PAISF: A \underline{P}artially \underline{A}daptive \underline{I}n-Order \underline{S}election \underline{F}unction for Routing in On-Chip, Inter-Chiplet, and Wafer-Scale Networks}

\author{
\ifnb
\IEEEauthorblockN{Patrick Iff}
\IEEEauthorblockA{ETH Zurich\\Switzerland\\
iffp@inf.ethz.ch\vspace{-2em}}
\and
\IEEEauthorblockN{Luca Oertig}
\IEEEauthorblockA{ETH Zurich\\Switzerland\\
loertig@student.ethz.ch\vspace{-2em}}
\and
\IEEEauthorblockN{Andrin Weber}
\IEEEauthorblockA{ETH Zurich\\Switzerland\\
andrwebe@student.ethz.ch\vspace{-2em}}
\and
\IEEEauthorblockN{Luca Benini}
\IEEEauthorblockA{ETH Zurich\\Switzerland\\
lbenini@iis.ee.ethz.ch\vspace{-2em}}
\and
\IEEEauthorblockN{Torsten Hoefler}
\IEEEauthorblockA{ETH Zurich\\Switzerland\\
htor@inf.ethz.ch\vspace{-2em}}
\else
\IEEEauthorblockN{Anonymous Authors}
\fi
}

\maketitle

\begin{abstract}
The growing number of compute elements per chip, along with advances such as chiplet systems and wafer-scale integration, drives the need for scalable and efficient routing algorithms for irregular topologies.
A routing algorithm consists of a routing function (RF) that ensures deadlock freedom and a selection function (SF) that avoids congestion and, in some cases, guarantees in-order packet delivery.
We propose two novel SFs for irregular topologies, ISF and PAISF, which ensure in-order delivery by hashing a packet's entropy value to determine its path.
Packets with the same entropy follow the same path and arrive in order.
Compared to a deterministic RF, a deterministic SF, and \textit{Multi}, three alternative in-order delivery strategies, (PA)ISF improves throughput by 15.2\% and reduces energy per packet and latency under load by 12.1\% and 11.3\%, at the cost of a zero-load latency increase by 3.3\% on average across workloads and topologies.

\end{abstract}


\section{Introduction}
\label{sec:intro}

The rising demand for compute power from ever larger \gls{ml} models drives growth in the number of compute elements per chip.
Since many \gls{ml} workloads are limited by data movement \cite{data-movement}, scalable and efficient \glspl{noc} are critical.
Although the \gls{m} topology has long been the de facto standard, more scalable topologies \cite{folded-torus,hypercube,flattened-butterfly} have recently emerged.
Furthermore, to sustain performance improvements despite the slowdown of Moore's law and the end of Dennard scaling, chiplet systems and \gls{wsi} have been adopted, giving rise to specialized topologies tailored for \glspl{ici} \cite{hexamesh,foldedhexatorus,double-butterfly,butterdonut,kite} and wafer-scale systems \cite{wsi-paper}.

A key component of any network is the routing algorithm, which determines how packets are forwarded from source to destination.
The routing algorithm is invoked whenever a packet arrives at a router and consists of two subroutines: the \gls{rf}, which identifies a set of next hops that guarantee deadlock and livelock freedom, and the \gls{sf}, which selects one of these hops to optimize performance by avoiding congestion.
While there is extensive work on \glspl{rf} \cite{tm,oetm,mr,padaptive} and \glspl{sf} \cite{pca,rca,dar,nop,shafiei} for mesh topologies, routing in recently proposed irregular \gls{noc}, \gls{ici}, and \gls{wsi} topologies remains less explored.

In this paper, we present two new \glspl{sf} for irregular topologies: \gls{isf} and \gls{paisf}, which adapts to current network congestion.
The key feature of both \glspl{sf} is that the sender can decide whether the packets of a given message need to be delivered \gls{io} or if delivering them \gls{ooo} is sufficient.
(PA)ISF ensures correct ordering independently of the underlying \gls{rf} by selecting the next hop through hashing an entropy value assigned to each packet, similar to the approach used in Ultra Ethernet \cite{ue}.
Assigning the same entropy value to all packets of a message ensures they follow the same path and remain in order, while assigning random entropy values allows packets to take different paths, exploiting path diversity and improving performance.
Ensuring in-order delivery is critical for many applications; if the network does not handle this, each endpoint requires a \gls{rob}, adding significant complexity and cost.
Moreover, if a \gls{rob} is full, packets must be dropped and retransmitted, potentially causing substantial performance degradation.

We compare our proposed \glspl{sf} against three approaches for ensuring in-order delivery.
First, a deterministic \gls{rf} that routes all packets of a source-destination pair along a single path, making an \gls{sf} unnecessary.
Second, a \gls{det} that always selects the same next hop, even when the \gls{rf} offers multiple alternatives.
Third, the \textit{Multi} \cite{multi, multi2} middle layer, which operates between \gls{rf} and \gls{sf}, identifies disjoint source-destination paths, and reorders packets at the destination router’s input buffer, avoiding a \gls{rob} at the endpoint while exploiting path diversity.
Compared to these baselines, our \glspl{sf} increase saturation throughput by 15.2\% and reduce energy per packet and latency under load by 12.1\% and 11.3\% on average, at the cost of a 3.3\% increase in zero-load latency.
In summary, the main contributions of this paper are:
\begin{itemize}
    \item We propose two novel in-order \glspl{sf} for irregular topologies, \gls{isf} and \gls{paisf}, which increase saturation throughput by 15.2\% on average compared to existing in-order delivery methods.
    \item We present the first comprehensive large-scale evaluation of \glspl{rf} and \glspl{sf} for irregular, state-of-the-art \gls{noc}, \gls{ici}, and \gls{wsi} topologies, covering multiple synthetic and real-world workloads and including ablation studies on virtual channel count, in-order traffic ratio, message size, and other factors.
\end{itemize}

\section{Background on On-Chip Routing}
\label{sec:back}

An on-chip, inter-chiplet, or wafer-scale network consists of routers connected by channels.
Endpoints, such as CPUs, GPUs, or \gls{ml} accelerators, are attached to the routers.
To communicate, a source endpoint injects a message into the network, which is routed to the destination endpoint.
A message is composed of packets, which in turn consist of \glspl{flit}.
Depending on application requirements, the packets of a message may need to be delivered in-order (io) or can be delivered out-of-order (ooo).
We assume wormhole routing, where only the header flit contains routing information and the remaining flits follow the header along the same path without interleaving with other packets.
We also assume credit-based flow control, meaning a router can only send a flit if the next router has sufficient buffer space to receive it.
With credit-based flow control, adaptive \glspl{sf} can track the buffer occupancy of the next router through available credits.
We use networks with multiple \glspl{vc} per physical channel, giving each router one input buffer per \gls{vc}.
\glspl{vc} decouple buffer resources from channel resources, allowing blocked packets to be overtaken by others and improving performance.
Some \glspl{rf}, such as \gls{lash}, use \glspl{vc} to prevent deadlocks.
An important implication is that packets on the same physical channel but in different \glspl{vc} can overtake each other, potentially causing out-of-order delivery.

\begin{figure*}[h]
\centering
\captionsetup{justification=centering}
\includegraphics[width=1.0\textwidth]{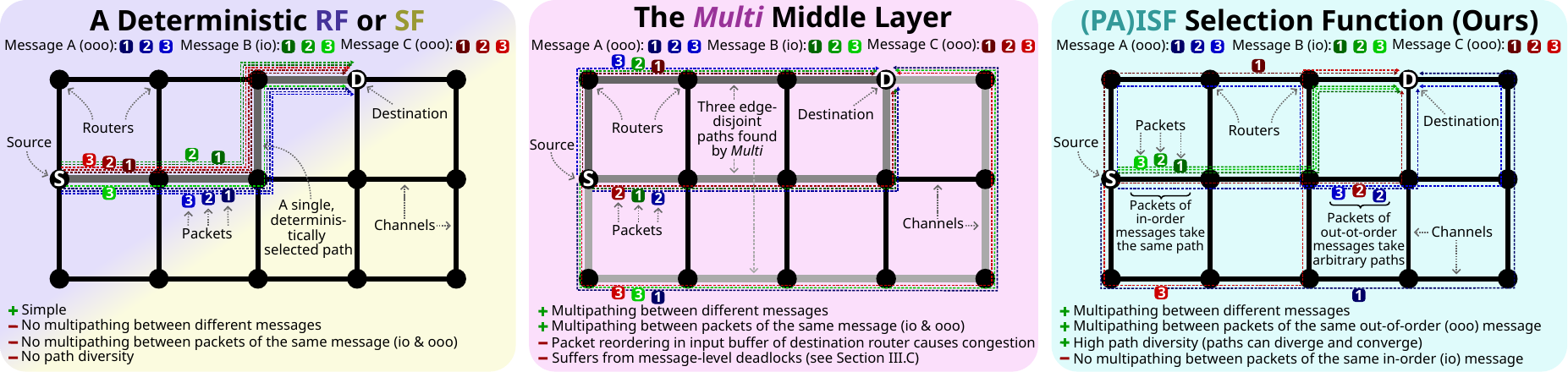}
\vspace{-1.5em}
\caption{{(\textsection \ref{sec:idea}) Overview of strategies to achieve in-order (io) packet delivery.}}
\label{fig:overview-ios}
\vspace{-2.0em}
\end{figure*}

\section{Related Work}
\label{sec:rl}

\subsection{Routing Functions}
\label{sec:rl-rf}

Routing functions (RFs) guarantee deadlock freedom by ensuring that the \gls{cdg} is acyclic \cite{cdg-dally,cdg-duato-1}.
The \gls{cdg} is a directed graph where vertices represent (virtual) channels and directed edges indicate that the \gls{rf} can forward packets from one channel to another.
In 2D mesh networks, this is achieved by restricting specific turns, for example using the Turn Model \cite{tm} or the Odd-Even Turn Model \cite{oetm}, by constraining the use of \glspl{vc} as in MESH\_ROUTE \cite{mr}, or by combining both approaches, as in P-Adaptive \cite{padaptive}.

\glspl{rf} for irregular topologies, where no regular structure or clear notion of turns exists, such as Up*/Down* \cite{updown}, Prefix routing \cite{prefix-routing}, Left/Right \cite{lturn}, \gls{scb} \cite{scb}, and L-Turn \cite{lturn}, employ more sophisticated mechanisms to break cycles in the \gls{cdg}.
\gls{lash} \cite{lash}, LASH-TOR \cite{lash-tor}, and DFSSSP \cite{dfsssp} achieve deadlock freedom by partitioning the network into routing layers, each using its own set of \glspl{vc}.
The number of required \glspl{vc} in \gls{lash} and DFSSSP depends on network size and topology, whereas LASH-TOR provides a fallback layer on the last \gls{vc}, using Up*/Down* routing to prevent deadlocks.
As shown in \Cref{tab:routing-functions}, most \glspl{rf} do not guarantee in-order delivery in irregular topologies with multiple \glspl{vc}.

\begin{table}[h]
\vspace{-0.5em}
\footnotesize
\setlength{\tabcolsep}{6.25pt}
\centering
\captionsetup{justification=centering}
\caption{\setstretch{0.87} \textbf{(\textsection \ref{sec:rl-rf}) Overview of routing functions}.\\
	\rone~Only guarantees in-order delivery in networks without \glspl{vc}.\\
	\rtwo~Guarantees in-order delivery if \# VCs equals \# routing layers. \\
	\rthree~Uses non-minimal paths if it needs to resort to a fallback layer.\\
	\rfour~Might resort to Up*/Down* routing in networks without \glspl{vc}.
}
\vspace{-0.5em}
\begin{tabular}{lccccc}
\toprule
\textbf{\makecell[l]{Name}}
& \rotatebox[origin=l]{90}{\textbf{\makecell[l]{Supports\\irregular\\topologies}}}
& \rotatebox[origin=l]{90}{\textbf{\makecell[l]{Provides multiple\\physical paths}}}
& \rotatebox[origin=l]{90}{\textbf{\makecell[l]{Includes non-\\minimal paths}}}
& \rotatebox[origin=l]{90}{\textbf{\makecell[l]{Supports\\networks without\\virtual channels}}}
& \rotatebox[origin=l]{90}{\textbf{\makecell[l]{Guarantees\\in-order delivery\\(Deterministic)}}}
\\
\midrule
XY 								& \xmark & \xmark & \xmark & \cmark & \rone	 \\
SCB \cite{scb}  				& \cmark & \cmark & \cmark & \cmark & \xmark \\
Up*/Down* \cite{updown} 		& \cmark & \cmark & \cmark & \cmark & \xmark \\
Prefix \cite{prefix-routing} 	& \cmark & \xmark & \cmark & \cmark & \rone	 \\
Left/Right \cite{lturn} 		& \cmark & \cmark & \cmark & \cmark & \xmark \\
L-Turn \cite{lturn} 			& \cmark & \cmark & \cmark & \cmark & \xmark \\
LASH \cite{lash} 				& \cmark & \xmark & \xmark & \xmark & \rtwo	 \\
LASH-TOR \cite{lash-tor} 		& \cmark & \xmark & \rthree& \rfour & \rtwo	 \\
DFSSSP \cite{dfsssp} 			& \cmark & \xmark & \xmark & \xmark & \rtwo	 \\
\bottomrule
\end{tabular}
\label{tab:routing-functions}
\vspace{-1.0em}
\end{table}

\subsection{Selection Functions}
\label{sec:rl-sf}

So far, \glspl{sf} have primarily been proposed in the context of mesh networks, focusing on congestion avoidance rather than in-order delivery.
PCA \cite{pca} and the \gls{sf} proposed by Shafiei et al. \cite{shafiei} use congestion information from neighboring routers to guide decisions.
NoP \cite{nop} and RCA \cite{rca} gather congestion information from a larger network region, while DAR \cite{dar} leverages global congestion information.
However, none of these \glspl{sf} support in-order delivery.

A simple way to achieve in-order delivery is to use a deterministic selection function (DET), which always selects the first (channel, \gls{vc}) pair from the candidate list returned by the \gls{rf} (assuming this list is always ordered in the same way).
As shown in \Cref{fig:overview-ios} (left), \gls{det} routes all packets of a source-destination pair along the same path, guaranteeing in-order delivery but failing to exploit path diversity and additional input buffers offered by \glspl{vc}.

\subsection{The Multi Middle Layer for In-Order Delivery}
\label{sec:rl-multi}

\begin{figure}[h]
\vspace{-1.0em}
\centering
\captionsetup{justification=centering}
\includegraphics[width=1.0\columnwidth]{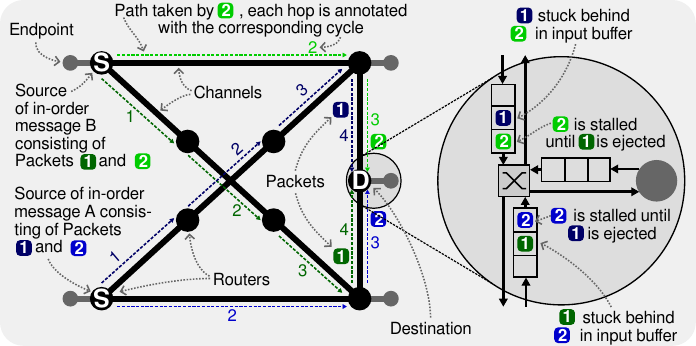}
\vspace{-1.75em}
	\caption{{(\textsection \ref{sec:rl-multi}) Example of a message-level deadlock in \textit{Multi}.}}
\label{fig:multi-dl}
\vspace{-0.5em}
\end{figure}

\textit{Multi} \cite{multi,multi2} can be placed between the \gls{rf} and \gls{sf} to guarantee in-order delivery while exploiting path diversity.
Using only the deadlock-free paths provided by the \gls{rf}, \textit{Multi} computes a maximal set of channel- and \gls{vc}-disjoint paths between source and destination, as shown in \Cref{fig:overview-ios} (center).
At the source, any \gls{sf} can be used to select one of these paths.
Packets remain ordered along each path.
To ensure global ordering, the destination router dequeues only the packet with the next expected sequence number from the input buffer and stalls the others.
Because packets are ordered within each path, the expected packet eventually reaches the head of its input buffer.
A drawback of \textit{Multi} is that it can cause message-level deadlocks, as illustrated in \Cref{fig:multi-dl}.
Such a deadlock occurs when two messages A and B, consisting of packets \pkgaone, \pkgatwo~and \pkgbone, \pkgbtwo~respectively, are sent to the same destination, and \pkgatwo~blocks \pkgbone~while \pkgbtwo~blocks \pkgaone.
This issue can likely be mitigated using shared buffers or strategies based on \glspl{vc} or multiplanes.

\begin{figure*}[h]
\centering
\captionsetup{justification=centering}
\includegraphics[width=1.0\textwidth]{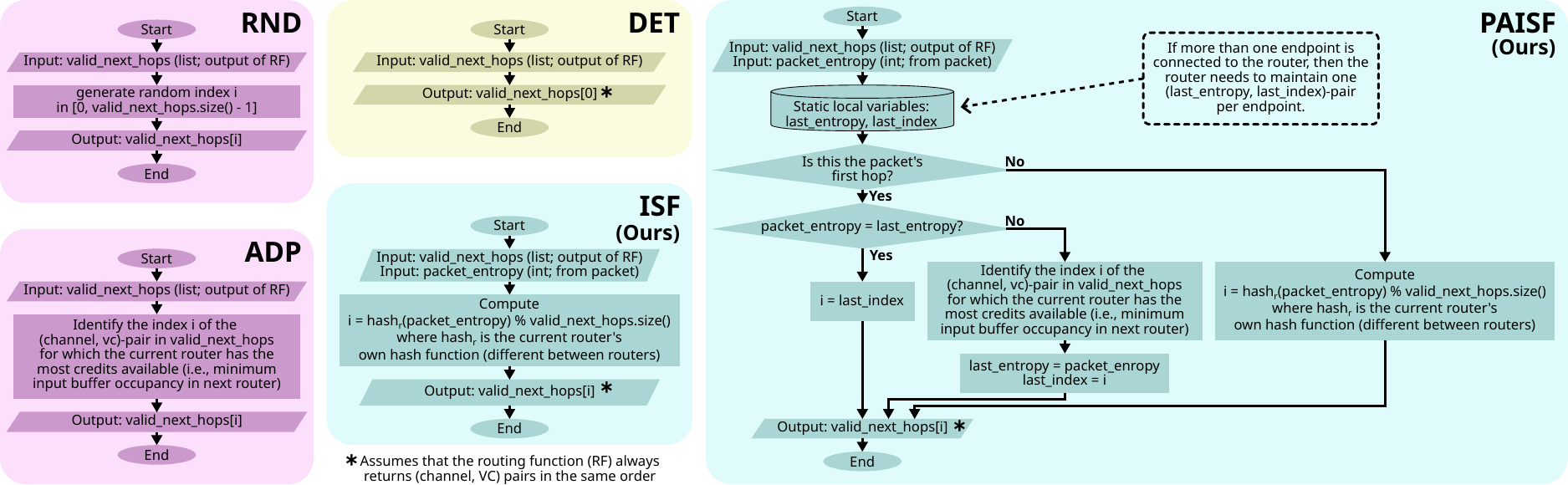}
\vspace{-1.5em}
\caption{{(\textsection \ref{sec:idea}) Overview of existing selection functions (RND, ADP, DET) and our proposed selection functions (ISF and PAISF).}}
\label{fig:overview-sf}
\vspace{-2.0em}
\end{figure*}

The authors of \textit{Multi} also propose a selection function that leverages an offline \gls{ilp} to compute the optimal path selection for each source-destination pair using traffic information available at design time.
Since we assume traffic is unknown at design time, we instead use two online selection functions for \textit{Multi}: \gls{rnd}, which distributes packets uniformly at random across available paths, and \gls{adp}, which selects the path whose next hop has the lowest buffer occupancy.
Both \glspl{sf} are illustrated in \Cref{fig:overview-sf} (left).

\section{The ISF and PAISF Selection Functions}
\label{sec:idea}

The \gls{sf} selects one of the (channel, \gls{vc}) pairs returned by the \gls{rf}.
Its goal is to maximize performance by exploiting path diversity through multiple physical paths and by adapting to current network congestion.
In systems requiring in-order delivery, the \gls{sf} can also ensure that packets of the same message arrive in order.

\glsreset{det} \glsreset{rnd} \glsreset{adp} \glsreset{isf} \glsreset{paisf}

The \gls{det} ensures in-order delivery by always selecting the same (channel, \gls{vc}) pair for a given destination (packets cannot be reordered if they follow the same path and use the same \gls{vc}), but it neither exploits path diversity nor adapts to network conditions.
The \gls{rnd} exploits path diversity by randomly selecting a (channel, \gls{vc}) pair, yet it is non-adaptive and does not guarantee in-order delivery.
The \gls{adp} exploits path diversity and adapts to congestion by choosing the (channel, \gls{vc}) pair whose next router has the lowest input buffer occupancy, but it also does not ensure in-order delivery.
We therefore propose two novel \glspl{sf}, \gls{isf} and \gls{paisf}, which both provide fine-grained ordering control by allowing applications to specify per-message ordering requirements.
Both exploit path diversity, and \gls{paisf} further adapts to congestion by selecting the first hop of a path based on congestion information.
Detailed descriptions of all five \glspl{sf} are given in \Cref{fig:overview-sf}, and their properties are compared in \Cref{tab:selection-functions}.

\begin{table}[h]
\vspace{-0.75em}
\footnotesize
\setlength{\tabcolsep}{10pt}
\centering
\captionsetup{justification=centering}
\caption{\textbf{(\textsection \ref{sec:idea}) Overview of selection functions}.}
\vspace{-0.75em}
\begin{tabular}{lccc}
\toprule
\textbf{\makecell[l]{Name}}
& \rotatebox[origin=c]{0}{\textbf{\makecell[c]{Exploits\\path\\diversity}}}
& \rotatebox[origin=c]{0}{\textbf{\makecell[c]{Adapts to\\network\\congestion}}}
& \rotatebox[origin=c]{0}{\textbf{\makecell[c]{Guarantees\\in-order\\delivery}}}
\\
\midrule
Deterministic (DET) 			& \xmark & \xmark & \cmark \\
Random (RND) 					& \cmark & \xmark & \xmark \\
Adaptive (ADP) 					& \cmark & \cmark & \xmark \\
\midrule
\textbf{ISF} (Ours) 			& \cmark & \xmark & \cmark \\
\textbf{PAISF} (Ours) 			& \cmark & \cmark & \cmark \\
\bottomrule
\end{tabular}
\label{tab:selection-functions}
\vspace{-0.5em}
\end{table}

\glsreset{io} \glsreset{ooo}

\underline{The key idea of \textbf{\gls{isf}}} is that each router $r$ implements its own hash function $h_r$ and each packet carries a packet entropy $e_p$ in its header.
When a packet arrives, the router calls its \gls{rf} or looks up the precomputed routing table to obtain the list $\text{\texttt{valid\_next\_hops}}$ of (channel, \gls{vc}) pairs that forward the packet toward its destination while guaranteeing deadlock and livelock freedom.
\gls{isf} then selects the pair at index $h_r(e_p) \in [0, |\text{\texttt{valid\_next\_hops}}|)$, assuming that the list of valid next hops is ordered in a deterministic way.
All packets with the same $e_p$ follow the same path and use the same \gls{vc}, ensuring in-order delivery, while packets with different $e_p$ may take different paths, exploiting path diversity.
For messages without in-order requirements, the packetization layer assigns random entropy values to each packet, allowing full path diversity.
For messages that require in-order delivery, the same entropy is assigned to all packets of a message, and different entropies to different messages, enabling \gls{isf} to maintain in-order delivery within each message while exploiting path diversity across messages.
\Cref{fig:overview-ios} (right) visualizes how \gls{isf} routes \gls{io} and \gls{ooo} messages.

As we show in \Cref{sec:eval-entropy}, using $b_e = 3$ entropy bits per packet is sufficient to achieve good performance in typical on-chip networks while keeping packet header overhead minimal.
For a network with $n$ routers, the hash function $h_r$ can be efficiently implemented with an internal register width of $w = \max(b_e, \lceil \log_2(n) \rceil) + 2$ bits.
The function first computes $z = r \oplus e_p$ and multiplies $z$ by a constant $c$ to mix the bits.
Finally, it maps the result to $[0, |\textit{\texttt{valid\_next\_hops}}|)$ using a multiplication followed by a right shift, where $|\textit{\texttt{valid\_next\_hops}}|$ is upper-bounded by the router radix times the number of \glspl{vc} per channel, both typically around four.
The resulting hash function is
\begin{equation*} 
	h_r(e_p) = \Big( \underbrace{(r \oplus e_p) \cdot c}_{\substack{\text{truncated to $w$ bits} \\ \text{by register width}}} \cdot |\textit{\texttt{valid\_next\_hops}}| \Big) \cdot \underbrace{2^{-w}}_{\text{right shift}},
\end{equation*}
and requires only $b_e$ XOR gates, a few shifted additions for the multiplication by the constant $c$, and a small $w \times \lceil \log_2 |\textit{\texttt{valid\_next\_hops}}| \rceil$ multiplier.
For $n = 64$ routers and up to 16 valid next hops, this amounts to roughly 400 gate equivalents per input port, which is negligible compared to the router's input buffers and crossbar.

\textbf{\gls{paisf}} follows \gls{isf} by using the same hash mechanism to ensure in-order delivery.
Additionally, at the injection router, \gls{paisf} selects the first hop based on congestion, choosing the (channel, \gls{vc}) pair whose next router input buffer has the lowest occupancy (known through credit-based flow control).
To maintain in-order delivery, each router stores, for every attached endpoint, the last packet entropy received and the corresponding (channel, \gls{vc}) pair selected.
When a new packet arrives from that endpoint, the router checks its entropy: if it matches the last received value, it uses the same (channel, \gls{vc}) pair; otherwise, it selects a new pair based on congestion and updates the stored entropy and (channel, \gls{vc}) pair.
This ensures in-order delivery for messages that require it while performing adaptive load-balancing across messages and across packets of out-of-order messages.
Compared to \gls{isf}, \gls{paisf} requires two registers of extra storage per endpoint and a more complex selection process, as illustrated in \Cref{fig:overview-sf} (right).

\section{Evaluation}
\label{sec:eval}

\glsreset{det}

We evaluate \gls{isf} and \gls{paisf} against three existing in-order delivery methods: the deterministic LASH-TOR (TOR) \cite{lash-tor} \gls{rf}, using as many layers as available \glspl{vc} and balancing paths across layers; the \gls{det}; and the \textit{Multi} \cite{multi} middle layer, evaluated with both \gls{rnd} and \gls{adp} \glspl{sf}.
\textit{Multi}, \gls{det}, and (PA)ISF are evaluated across six \glspl{rf}: XY, SCB \cite{scb}, Up*/Down* (UD) \cite{updown}, L-Turn (LT) \cite{lturn}, Left/Right (LR) \cite{lturn}, and Prefix routing (P) \cite{prefix-routing}.
Since XY routing is only applicable to mesh topologies, we report \texttt{N/A} for XY on non-mesh topologies.  
We exclude LASH \cite{lash} and DFSSSP \cite{dfsssp} from our evaluation because, on the investigated topologies (see \Cref{sec:eval-topologies}), they require up to six and eight \glspl{vc}, which exceeds the four \glspl{vc} used in our setup (see \Cref{sec:eval-params}).

\subsection{Experiment Methodology and Evaluation Metrics}
\label{sec:eval-setup}

We use the cycle-accurate BookSim2 \cite{booksim} \gls{noc} simulator, in which we implement the \glspl{rf} listed in \Cref{tab:routing-functions} and the \glspl{sf} listed in \Cref{tab:selection-functions}. 
Through flit-level simulations in BookSim2, we measure the zero-load latency (in cycles) and saturation throughput (in flits per endpoint per cycle) for various topologies and workloads. 
The energy consumption per transmitted packet is estimated from two components: router power and link power.
Router power is computed using the Orion3 \cite{orion} model, scaled from 45\,nm to 7\,nm with DeepScaleTool \cite{deepscaletool}.
Link power is computed by counting the number of flits sent over each link, multiplied by the link length, and by an energy-per-flit-per-mm factor.
BookSim2 simulations are run three times.
We report arithmetic means for latency, harmonic means for throughput, and geometric means for relative performance, following scientific benchmarking guidelines \cite{th-bench-rules}.
Nonparametric 95\% \glspl{ci} for each metric are computed via bootstrap resampling. The mean relative half-width, defined as half the \gls{ci} width divided by the sample mean, is 15.80\% for throughput and 2.82\% for latency.
If message-level deadlocks in \textit{Multi} (see \Cref{sec:rl-multi}) cannot be resolved after ten retries with different random seeds, we report \texttt{DL} in the results.
In heatmap plots, we highlight the best-performing \gls{rf}-\gls{sf} combination for each topology-traffic pair in bold, and underline the second-best combination. 

\subsection{Simulation Parameters}
\label{sec:eval-params}

Unless stated otherwise, messages consist of 16 packets of four flits each, and 50\% of messages require in-order delivery.
The network operates at $1$ GHz and we use $64$-bit flits for \glspl{noc} and \glspl{ici}, and $16$-kbit flits for \gls{wsi}, resulting in $2$ TB/s bandwidth as in Tesla's Dojo \cite{dojo}.
We assume a router latency of three cycles, four \glspl{vc} per physical channel, and \gls{vc} buffers of 8, 16, or 32 flits for \gls{noc}, \gls{ici}, and \gls{wsi}, respectively, as longer links require more buffering to maintain high performance.
Link energy is 0.1\,pJ/bit/mm \cite{energy-ref}, and link latency is one cycle per 2\,mm, modeling a register buffer every 2\,mm, with four-cycle \gls{phy} latency for \glspl{ici}.

\subsection{Evaluated Workloads}
\label{sec:eval-workloads}

We consider three synthetic traffic patterns: uniform (\texttt{U}), representing all-to-all communication as found in \gls{moe} training \cite{uniform-ref}; random permutation (\texttt{RP}), capturing shuffle-like communication such as FFT or sorting \cite{randperm-ref}; and asymmetric (\texttt{A}), modeling producer-consumer scenarios.
Beyond synthetic traffic, we also evaluate CPU and GPU workloads using traces truncated to 100k events, since simulating millions of events in BookSim2 would be prohibitively time-consuming.
For CPUs, we employ Netrace \cite{netrace} with the \textit{blackscholes} trace, which represents a PDE-based financial workload computing option prices via the Black-Scholes equation.
For GPUs, we use the ATLAHS \cite{atlahs} toolchain to generate GOAL \cite{goal} traces from training the \textit{Llama-7B} \cite{llama} large language model.

\begin{table}[h]
\vspace{-0.5em}
\scriptsize
\setlength{\tabcolsep}{5.5pt}
\centering
\captionsetup{justification=centering}
\caption{\textbf{(\textsection \ref{sec:eval-topologies}) Overview of investigated network topologies}.}
\vspace{-0.5em}
\begin{tabular}{llccccc}
\toprule
\rotatebox[origin=l]{0}{\textbf{\makecell[l]{Domain}}}
& \textbf{\makecell[l]{Topology}}
	& \rotatebox[origin=l]{90}{\textbf{\makecell[l]{\hspace{-0.6em}Routers}}}
& \rotatebox[origin=l]{90}{\textbf{\makecell[l]{\hspace{-0.6em}Endpoints}}}
& \rotatebox[origin=l]{90}{\textbf{\makecell[l]{\hspace{-0.6em}Direct\\\hspace{-0.6em}Network}}}
& \rotatebox[origin=l]{90}{\textbf{\makecell[l]{\hspace{-0.6em}Network\\\hspace{-0.6em}Diameter}}}
& \rotatebox[origin=l]{90}{\textbf{\makecell[l]{\hspace{-0.6em}Router\\\hspace{-0.6em}Radix}}}
\\
\midrule
\multirow{5}{*}{\rotatebox[origin=c]{0}{NoC}}
	& Mesh (M)												& 64  & 64  & \cmark  & 14 	& 2-4  \\
	& Torus (T)												& 64  & 64  & \cmark  & 8 	& 4    \\
	& Folded Torus (FT) \cite{folded-torus}					& 64  & 64  & \cmark  & 8	& 4    \\
	& Hypercube (HC) \cite{hypercube}						& 64  & 64  & \cmark  & 6 	& 6   \\
	& Flattened Butterfly (FBF) \cite{flattened-butterfly} 	& 64  & 64  & \cmark  & 2	& 14   \\
\midrule
\multirow{4}{*}{\rotatebox[origin=c]{0}{ICI-P}}
	& Mesh (M)												& 64  & 64  & \cmark  & 14	& 2-4  \\
	& Folded Torus (FT) \cite{folded-torus}					& 64  & 64  & \cmark  & 8 	& 4    \\
	& HexaMesh (HM) \cite{hexamesh}							& 61  & 61  & \cmark  & 8	& 3-6  \\
	& FoldedHexaTorus (FHT) \cite{foldedhexatorus}			& 61  & 61  & \cmark  & 5	& 6    \\
\midrule
\multirow{5}{*}{\rotatebox[origin=c]{0}{ICI-A}}
	& Mesh (M)												& 64  & 64  & \cmark  & 14 	& 2-4  \\
	& Double Butterfly (DBF) \cite{double-butterfly}		& 64  & 64  & \cmark  & 8	& 2-4  \\
	& ButterDonut (BD) \cite{butterdonut}					& 64  & 64  & \cmark  & 6	& 3-4  \\
	& Kite Small (KS) \cite{kite}							& 64  & 64  & \cmark  & 7	& 3-4  \\
	& Kite Large (KL) \cite{kite}							& 64  & 64  & \cmark  & 8	& 2-4  \\
\midrule
\multirow{4}{*}{\rotatebox[origin=c]{0}{WSI}}
	& Mesh-Like (ML) \cite{wsi-paper}						& 316 & 64  & \xmark  & 31	& 2-4  \\
	& Aligned (A) \cite{wsi-paper}							& 188 & 64  & \xmark  & 25	& 2-5  \\
	& Interleaved (I) \cite{wsi-paper}						& 188 & 64  & \xmark  & 25	& 1-5  \\
	& Rotated (R) \cite{wsi-paper}							& 318 & 66  & \xmark  & 17	& 2-7  \\
\bottomrule
\end{tabular}
\label{tab:topologies}
\vspace{-2em}
\end{table}

\subsection{Evaluated Network Topologies}
\label{sec:eval-topologies}

We evaluate \gls{isf} and \gls{paisf} on established and emerging topologies for \glspl{noc} \cite{folded-torus,hypercube,flattened-butterfly}, \gls{wsi} \cite{wsi-paper}, and \glspl{ici} \cite{kite,butterdonut,double-butterfly,hexamesh,foldedhexatorus}, distinguishing between passive-substrate (ICI-P) and active-interposer (ICI-A)  designs.
We select topologies with approximately $64$ endpoints and, except for \gls{wsi}, use one endpoint per router.
The evaluated \gls{wsi} topologies are indirect networks composed of a compute wafer and a separate interconnection wafer, the latter containing many routers but no endpoints.
Details of all topologies are provided in \Cref{tab:topologies}.

\subsection{Results for Networks-on-Chip}
\label{sec:eval-noc}

\begin{figure}[h]
\vspace{-1.0em}
\centering
\captionsetup{justification=centering}
\includegraphics[width=1.0\columnwidth]{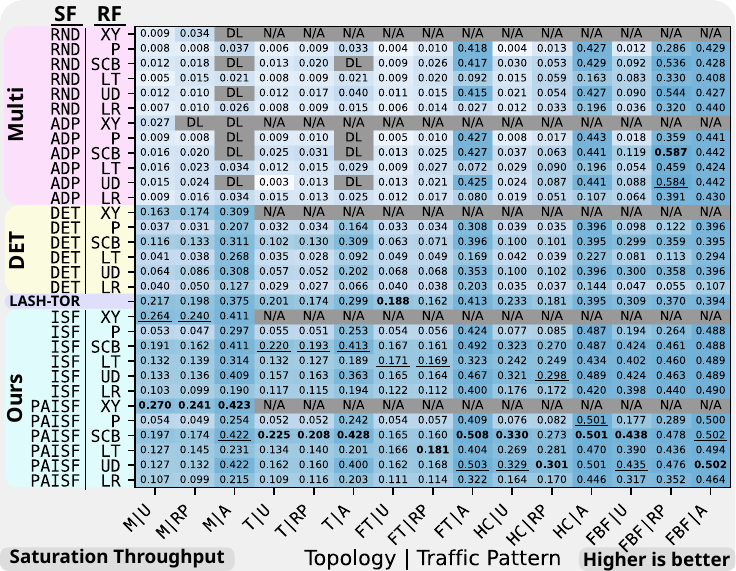}
\vspace{-1.5em}
	\caption{{(\textsection \ref{sec:eval-noc}) Saturation throughput on \gls{noc} topologies.}\\Traffic: \texttt{U}: Uniform, \texttt{RP}: Random Permutation, \texttt{A}: Asymmetric.}
\label{fig:topo-noc-tp-hm}
\vspace{-1.5em}
\end{figure}

\begin{figure}[h]
\centering
\captionsetup{justification=centering}
\includegraphics[width=1.0\columnwidth]{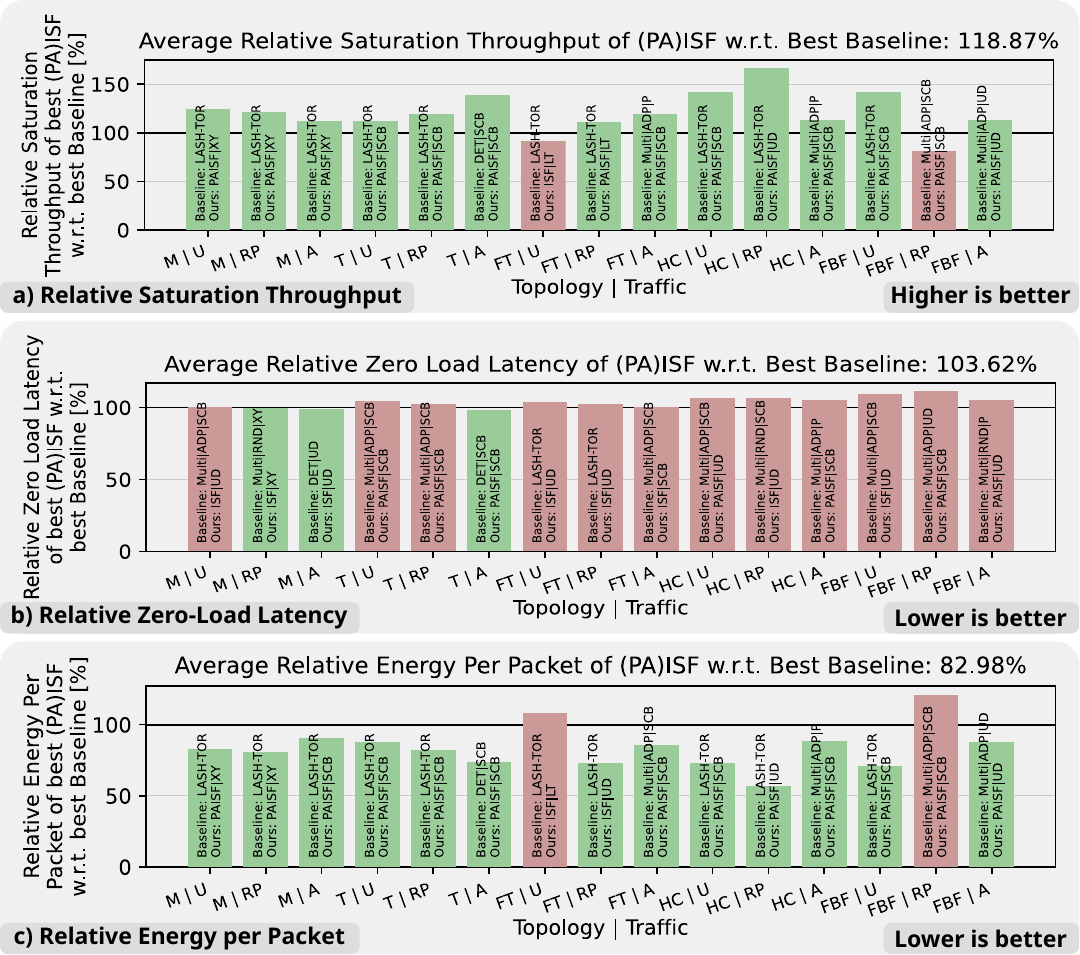}
\vspace{-1.5em}
\caption{{(\textsection \ref{sec:eval-noc}) Relative perf. of (PA)ISF w.r.t. best baseline (NoC).}}
\label{fig:topo-noc-impr}
\vspace{-2.0em}
\end{figure}

\Cref{fig:topo-noc-tp-hm} shows the saturation throughput of all \gls{rf}-\gls{sf} combinations for the evaluated \gls{noc} topologies, while \Cref{fig:topo-noc-impr} reports the relative saturation throughput, zero-load latency, and energy per packet of the best (PA)ISF configuration with respect to the best baseline (either \textit{Multi}, LASH-TOR, or \gls{det}) for each topology-traffic pair.
Detailed plots for latency and energy are available in our public repository\footnote{\label{fn:repo}\githuburl}.
\Cref{fig:paisf-vs-isf}a shows the relative throughput of \gls{paisf} with respect to \gls{isf}.

We make the following observations.
(1) (PA)ISF improves saturation throughput over \gls{det}, LASH-TOR, and \textit{Multi} in most cases, with an average gain of 18.9\%.
While \gls{det} and LASH-TOR do not exploit path diversity and thus achieve lower throughput, \textit{Multi} likely suffers from congestion due to packet reordering in the destination router’s input buffers.
(2) (PA)ISF increases zero-load latency by 3.6\% on average compared to the best baseline.
Since the average hop count is similar, the higher latency likely stems from increased arbitration overhead due to path divergence and reconvergence, not longer paths.
At an injection rate of 0.1, (PA)ISF decreases latency by 8.9\% on average (plot omitted due to space constraints), showing that its congestion-mitigation reduces latency under load.
(3) (PA)ISF reduces energy per packet in most cases, with an average improvement of 17.0\% over the best baselines.
(4) \Cref{fig:paisf-vs-isf}a shows that the additional adaptivity of \gls{paisf} provides only a minor throughput gain of 2.7\% over \gls{isf}.
(5) In mesh networks, XY routing often achieves higher throughput than routing functions for arbitrary topologies.
(6) Among routing functions for arbitrary topologies, LASH-TOR and SCB perform well across topologies, followed by Up*/Down*, L-Turn, Prefix routing, and Left/Right.

\begin{figure}[h]
\vspace{-1.0em}
\centering
\captionsetup{justification=centering}
\includegraphics[width=1.0\columnwidth]{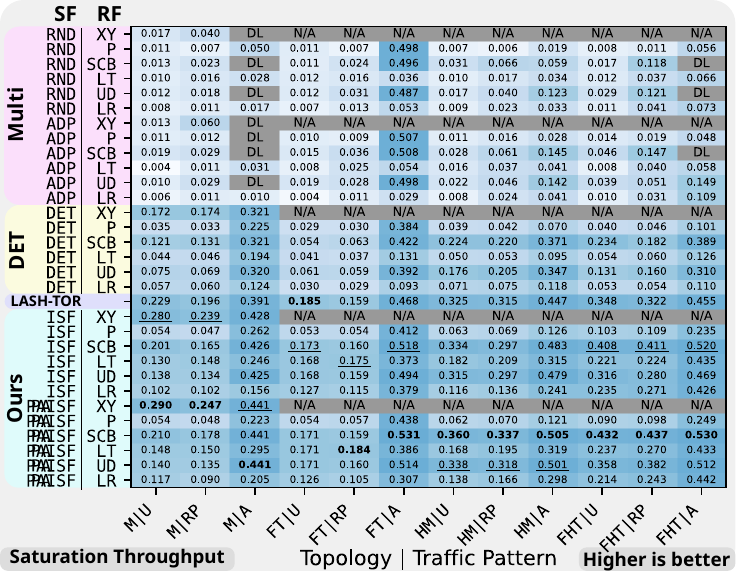}
\vspace{-1.75em}
\caption{{(\textsection \ref{sec:eval-pici}) Saturation throughput on ICI-P topologies.}}
\label{fig:topo-pici-tp-hm}
\end{figure}

\begin{figure}[h]
\centering
\captionsetup{justification=centering}
\includegraphics[width=1.0\columnwidth]{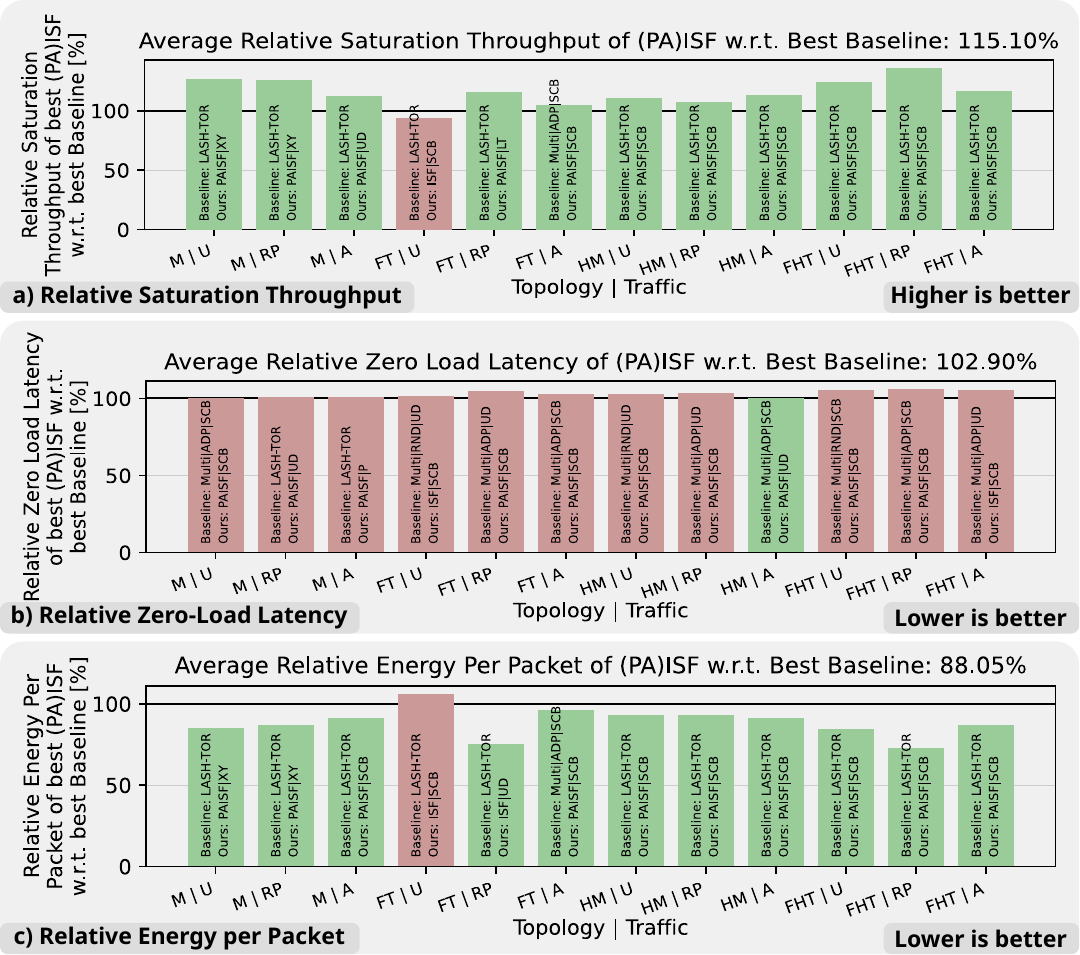}
\vspace{-1.5em}
\caption{{(\textsection \ref{sec:eval-pici}) Relative perf. of (PA)ISF w.r.t. best baseline (ICI-P).}}
\label{fig:topo-pici-impr}
\vspace{-1.0em}
\end{figure}

\subsection{Results for Inter-Chiplet Interconnects}
\label{sec:eval-ici}

Beyond the ubiquitous mesh, distinct topologies have been proposed for \glspl{ici} on passive substrates (ICI-P) and active interposers (ICI-A). We therefore evaluate (PA)ISF separately for both categories.

\subsubsection{ICI-P}
\label{sec:eval-pici}

\Cref{fig:topo-pici-tp-hm,fig:topo-pici-impr,fig:paisf-vs-isf}b report our results for ICI-P topologies.
The trends mirror those of \gls{noc}, with (PA)ISF increasing throughput by 15.1\% and reducing energy per packet by 11.9\% on average, at a 2.9\% latency increase without load and an 11.9\% latency reduction at an injection rate of 0.1.
PAISF’s adaptivity provides a 4.6\% average throughput gain over ISF.

\subsubsection{ICI-A}
\label{sec:eval-aici}

\begin{figure}[h]
\centering
\captionsetup{justification=centering}
\includegraphics[width=1.0\columnwidth]{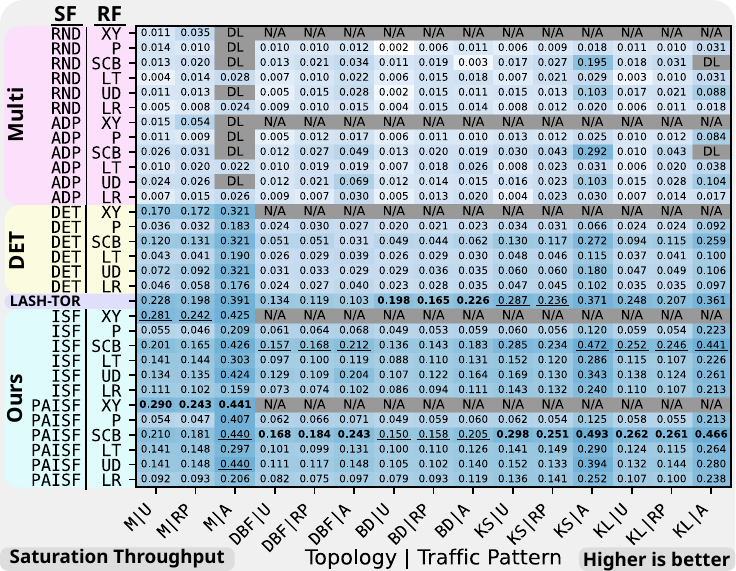}
\vspace{-1.5em}
\caption{{(\textsection \ref{sec:eval-aici}) Saturation throughput on ICI-A topologies.}}
\label{fig:topo-aici-tp-hm}
\end{figure}

\begin{figure*}[h]
\centering
\captionsetup{justification=centering}
\includegraphics[width=1.0\textwidth]{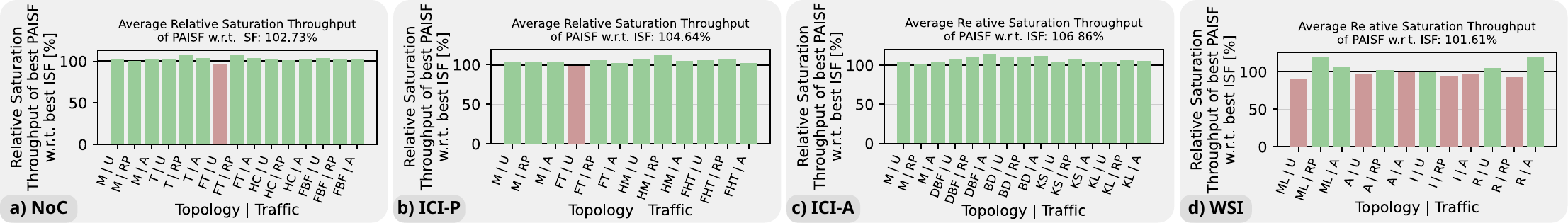}
\vspace{-1.5em}
\caption{{(\textsection \ref{sec:eval}) Relative saturation throughput of the best PAISF configuration w.r.t. the best ISF configuration.}}
\label{fig:paisf-vs-isf}
\vspace{-2.0em}
\end{figure*}

\begin{figure}[h]
\centering
\captionsetup{justification=centering}
\includegraphics[width=1.0\columnwidth]{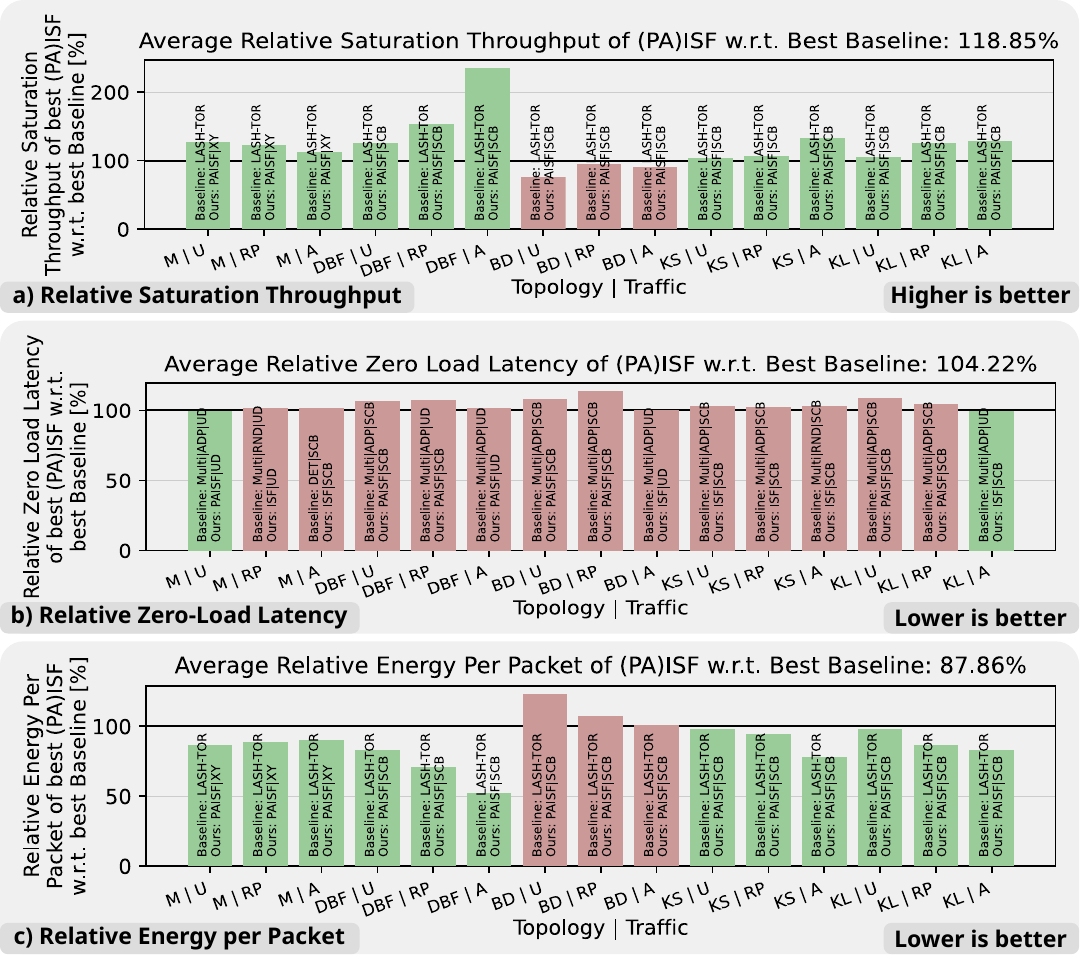}
\vspace{-1.5em}
\caption{{(\textsection \ref{sec:eval-aici}) Rel. perf. of (PA)ISF w.r.t. best baseline (ICI-A)}}
\label{fig:topo-aici-impr}
\vspace{-1.5em}
\end{figure}

\glsreset{scb}
\Cref{fig:topo-aici-tp-hm,fig:topo-aici-impr,fig:paisf-vs-isf}c show our results for ICI-A topologies.
Here, (PA)ISF achieves an average throughput gain of 18.9\% and reduces energy per packet by 12.1\%, with a 4.2\% latency increase at zero load and a 25.6\% latency reduction at an injection rate of 0.1 (plot omitted).
The \gls{scb} algorithm \cite{scb}, already strong in other domains, dominates in ICI-A and is only surpassed by XY routing on 2D meshes and by LASH-TOR \cite{lash-tor} on the ButterDonut topology.

\subsection{Results for Wafer-Scale Systems}
\label{sec:eval-wsi}

\begin{figure}[h]
\vspace{-1.0em}
\centering
\captionsetup{justification=centering}
\includegraphics[width=1.0\columnwidth]{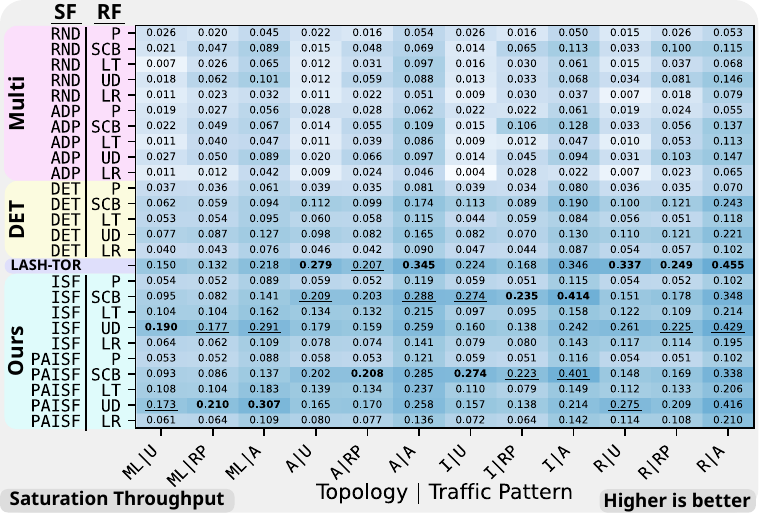}
\vspace{-1.5em}
\caption{{(\textsection \ref{sec:eval-wsi}) Saturation throughput on \gls{wsi} topologies.}}
\label{fig:topo-wsi-tp-hm}
\end{figure}

\begin{figure}[h]
\centering
\captionsetup{justification=centering}
\includegraphics[width=1.0\columnwidth]{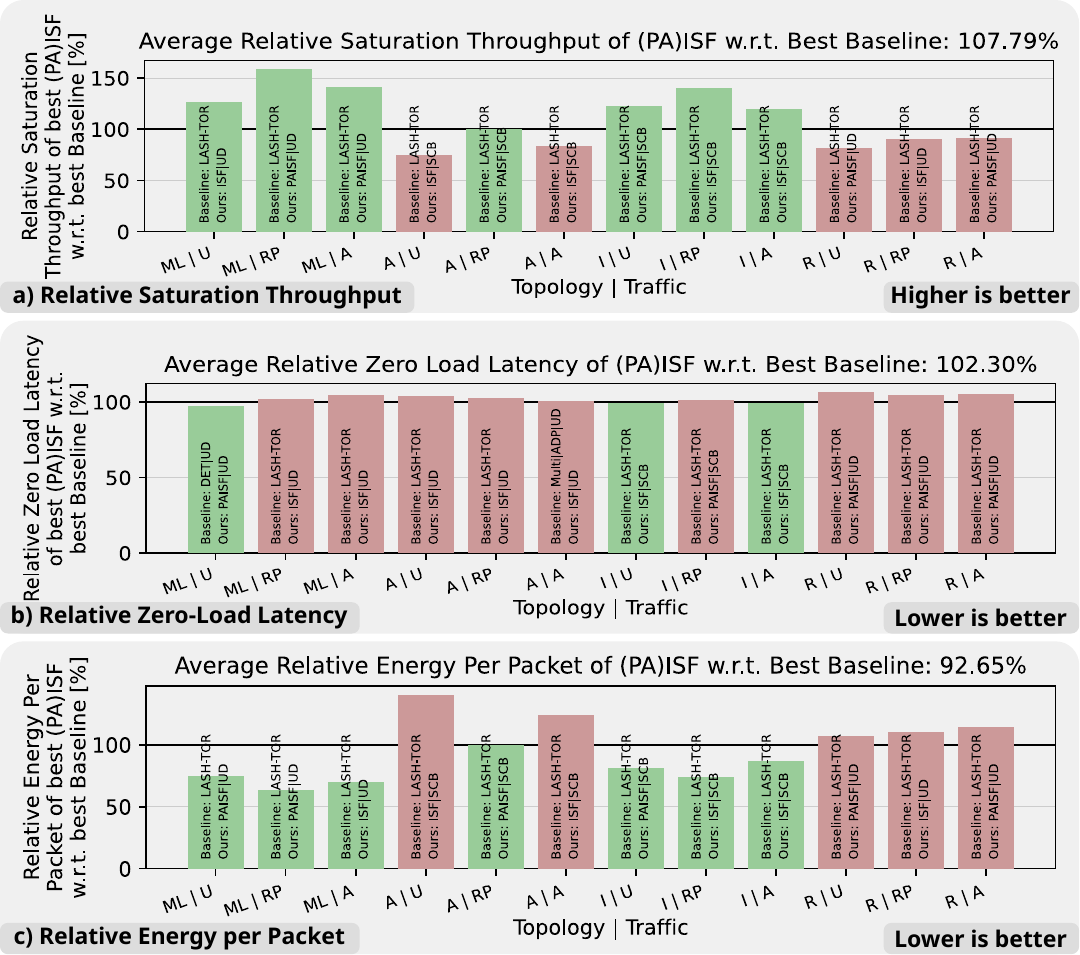}
\vspace{-1.5em}
\caption{{(\textsection \ref{sec:eval-wsi}) Relative perf. of (PA)ISF w.r.t. best baseline (WSI).}}
\label{fig:topo-wsi-impr}
\vspace{-1.5em}
\end{figure}

\Cref{fig:topo-wsi-tp-hm,fig:topo-wsi-impr,fig:paisf-vs-isf}d visualize our results on \gls{wsi} topologies.
On these topologies, with many more routers than endpoints due to a dedicated interconnection wafer, (PA)ISF achieves its lowest average throughput gain of 7.8\%, while reducing energy per packet by 7.3\%.
Zero-load latency and latency at an injection rate of 0.1 are increased by 2.3\% and 1.2\% respectively.
The throughput of \gls{isf} and \gls{paisf} is very close.
The limited benefits of (PA)ISF can likely be attributed to the fact that the average hop count in indirect \gls{wsi} topologies is roughly three times higher than in direct \gls{noc} and \gls{ici} topologies, making first-hop adaptivity insufficient.
Selecting the next hop with the lowest local congestion, as in \gls{paisf}, may not be sufficient to mitigate congestion in far downstream routers.

\subsection{Results for Traffic Traces}
\label{sec:eval-trace}

For trace-based evaluation, we use $64$ endpoints and exclude topologies that do not support this number.
The number of packets per message is set dynamically from the trace, and all messages are assumed to require in-order delivery.
Since throughput is determined by trace events rather than network capacity, we focus on latency. 

\subsubsection{CPU Workloads}
\label{sec:eval-trace-cpu}

\begin{figure}[h]
\centering
\captionsetup{justification=centering}
\includegraphics[width=1.0\columnwidth]{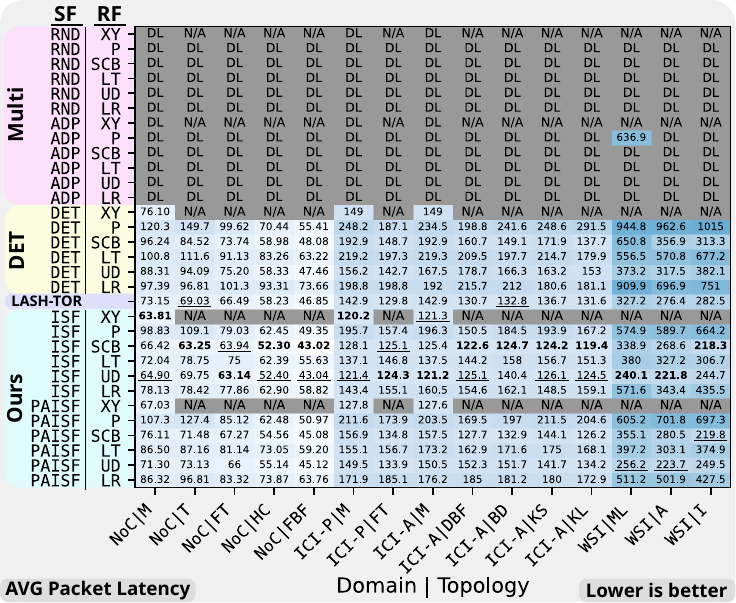}
\vspace{-1.5em}
\caption{{(\textsection \ref{sec:eval-trace-cpu}) Average packet latency for CPU trace.}}
\label{fig:trace-cpu-lat-hm}
\vspace{-1.0em}
\end{figure}

\begin{figure}[h]
\centering
\captionsetup{justification=centering}
\includegraphics[width=1.0\columnwidth]{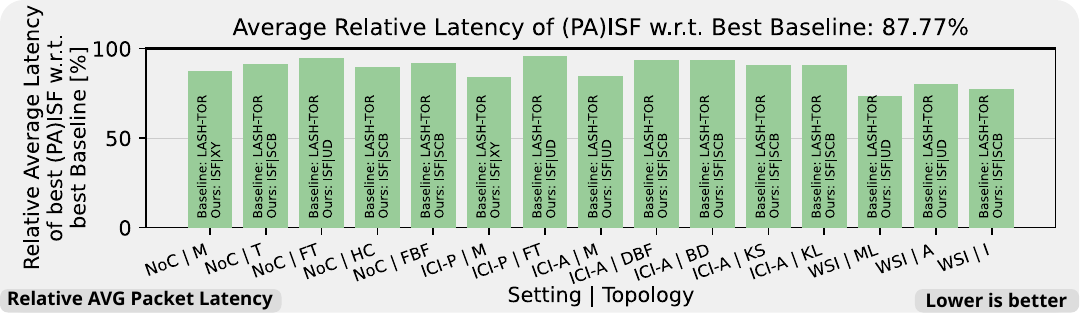}
\vspace{-1.5em}
\caption{{(\textsection \ref{sec:eval-trace-cpu}) Relative latency w.r.t. best baseline on CPU trace.}}
\label{fig:trace-cpu-lat}
\vspace{-1.5em}
\end{figure}

\Cref{fig:trace-cpu-lat-hm,fig:trace-cpu-lat} show the average packet latency and the improvement of the best (PA)ISF configuration over the best baseline for CPU traces.
(PA)ISF achieves an average latency reduction of 12.2\%.
This indicates that the higher zero-load latency observed under synthetic traffic is outweighed by reduced congestion-induced latency under realistic workloads.
In contrast to synthetic traffic, \textit{Multi} suffers from message-level deadlocks in most trace-based runs, likely due to more complex communication patterns.

\subsubsection{GPU Workloads}
\label{sec:eval-trace-gpu}

\begin{figure}[h]
\centering
\captionsetup{justification=centering}
\includegraphics[width=1.0\columnwidth]{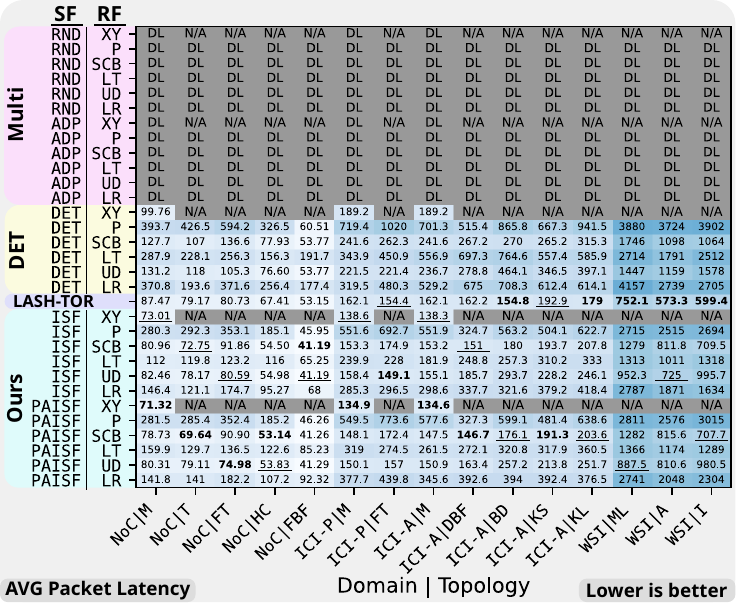}
\vspace{-1.5em}
\caption{{(\textsection \ref{sec:eval-trace-gpu}) Average packet latency for GPU trace.}}
\label{fig:trace-gpu-lat-hm}
\vspace{-1.0em}
\end{figure}

\begin{figure}[h]
\centering
\captionsetup{justification=centering}
\includegraphics[width=1.0\columnwidth]{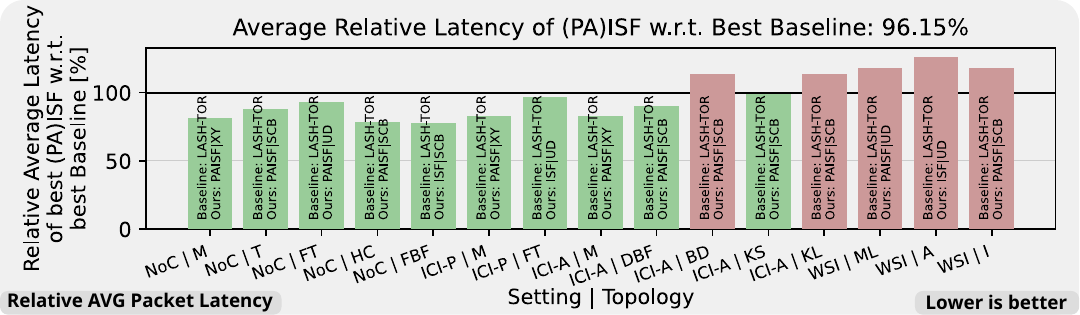}
\vspace{-1.5em}
\caption{{(\textsection \ref{sec:eval-trace-gpu}) Relative latency w.r.t. best baseline on GPU trace.}}
\label{fig:trace-gpu-lat}
\vspace{-1.5em}
\end{figure}

\Cref{fig:trace-gpu-lat-hm,fig:trace-gpu-lat} display our results for GPU traces.
Here, (PA)ISF and LASH-TOR achieve significantly lower latency than \textit{Multi} and \gls{det}, with (PA)ISF providing a 3.8\% average latency reduction over LASH-TOR.
The deadlock behavior of \textit{Multi} is similar to CPU traces.

\subsection{Ablation Studies}
\label{sec:eval-ablation}

To better understand (PA)ISF’s performance across scenarios, we conduct ablation studies varying seven key parameters.
We focus on uniform traffic and select one representative topology per domain (NoC, ICI-P, ICI-A, WSI).
We report saturation throughput as the primary metric, while latency and energy results are available in our public repository\footref{fn:repo}.
Each plot shows one line per \gls{rf}-\gls{sf} combination, colored by method (\textit{Multi}, LASH-TOR, \gls{det}, (PA)ISF), with the best-performing \gls{rf}-\gls{sf} combination per method highlighted in bold.
Error bars show 95\% \glspl{ci}.

\subsubsection{Varying Network Size}
\label{sec:eval-nwsize}

\begin{figure}[H]
\vspace{-1.0em}
\centering
\captionsetup{justification=centering}
\includegraphics[width=1.0\columnwidth]{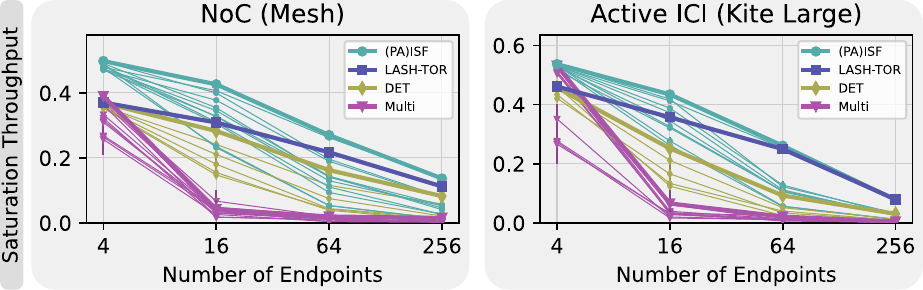}
\vspace{-1.75em}
\caption{{(\textsection \ref{sec:eval-nwsize}) Saturation throughput for varying network sizes.}}
\label{fig:nw-size-tp}
\vspace{-1.0em}
\centering
\end{figure}

We focus on \gls{noc} and ICI-A topologies, as they scale naturally to power-of-two sizes, and report results for $16$ to $256$ endpoints in \Cref{fig:nw-size-tp}. 
As the network grows, throughput per endpoint decreases as expected.
(PA)ISF performs best, with LASH-TOR matching its performance on large ICI-A topologies.
\textit{Multi}'s performance degrades significantly as the network size increases.

\subsubsection{Varying the In-Order Ratio}
\label{sec:eval-ior}

\begin{figure*}[t]
\centering
\captionsetup{justification=centering}
\includegraphics[width=1.0\textwidth]{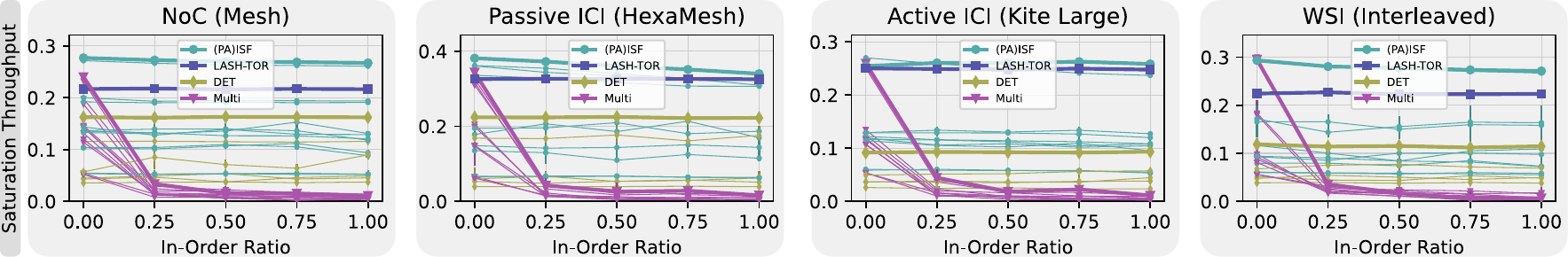}
\vspace{-1.5em}
\caption{{(\textsection \ref{sec:eval-ior}) Saturation throughput for varying ratios of in-order vs total messages.}}
\label{fig:ior-tp}
\vspace{-0.25em}
\end{figure*}

\Cref{fig:ior-tp} shows that the throughput of (PA)ISF slightly decreases as the in-order ratio increases. 
This is expected, as all packets of in-order messages follow the same path, limiting path diversity.
While \gls{det} and LASH-TOR are insensitive to the in-order ratio, \textit{Multi}'s performance drops significantly as the in-order ratio reaches 0.25 and remains consistently low thereafter.

\subsubsection{Varying Message Size}
\label{sec:eval-msgsize}

\begin{figure*}[t]
\centering
\captionsetup{justification=centering}
\includegraphics[width=1.0\textwidth]{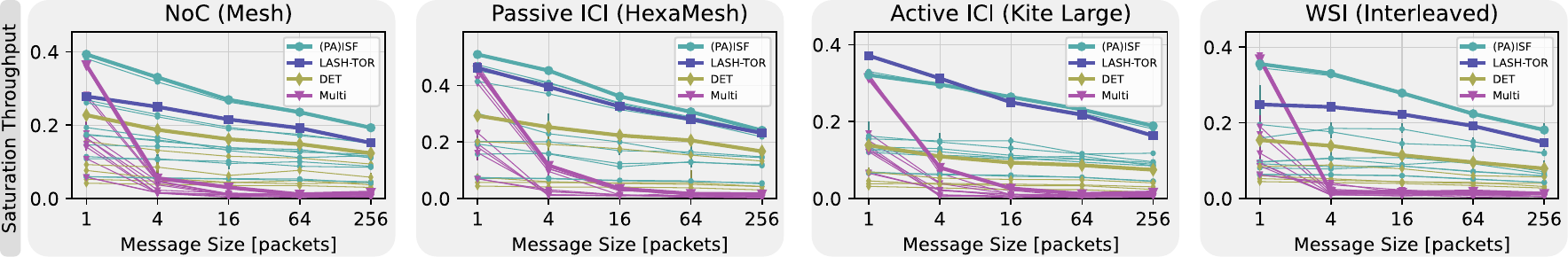}
\vspace{-1.5em}
\caption{{(\textsection \ref{sec:eval-msgsize}) Saturation throughput for varying message sizes.}}
\label{fig:msg-size-tp}
\vspace{-0.25em}
\end{figure*}

In \Cref{fig:msg-size-tp}, the throughput of all methods decreases with increasing message size.
\textit{Multi}, which matches (PA)ISF for single-packet messages where no reordering occurs, experiences a significant throughput drop as message size increases.
This indicates that \textit{Multi}'s primary performance bottleneck is packet reordering in the destination router's input buffers.

\subsubsection{Analysis under Link and Router Faults}
\label{sec:eval-fault}

\begin{figure*}[t]
\centering
\captionsetup{justification=centering}
\includegraphics[width=1.0\textwidth]{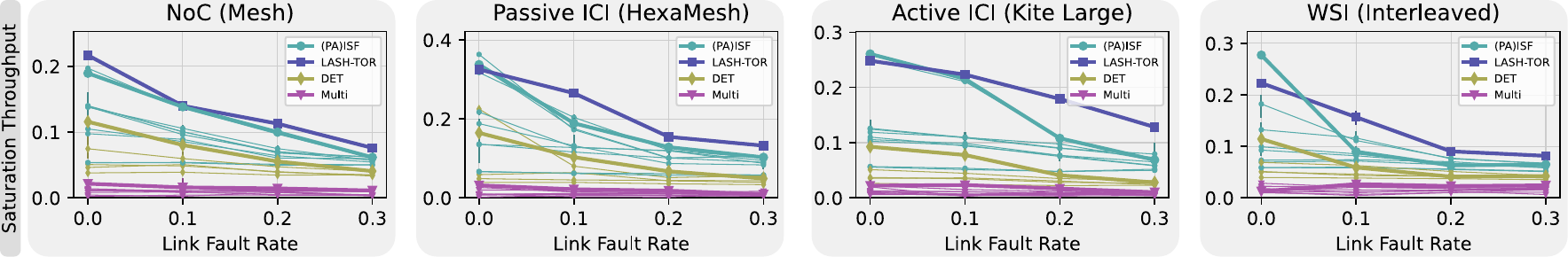}
\vspace{-1.5em}
\caption{{(\textsection \ref{sec:eval-fault}) Saturation throughput for varying link fault rates.}}
\label{fig:linkfault-tp}
\vspace{-0.25em}
\end{figure*}

\begin{figure*}[t]
\centering
\captionsetup{justification=centering}
\includegraphics[width=1.0\textwidth]{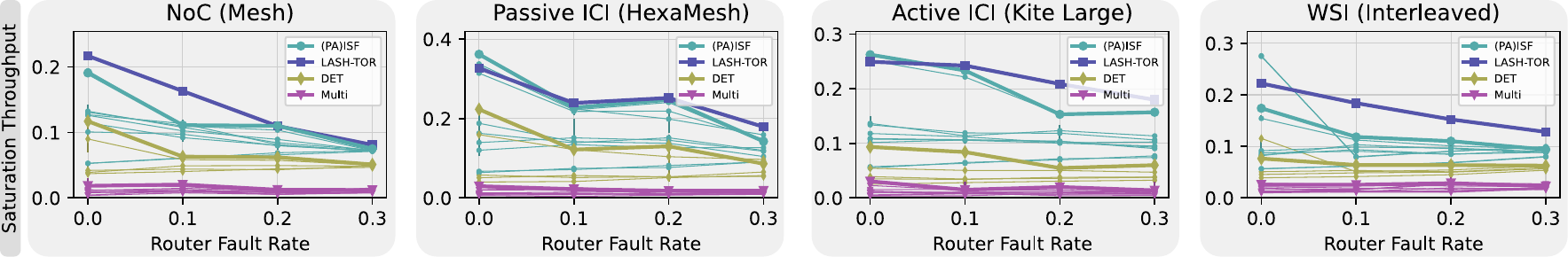}
\vspace{-1.5em}
\caption{{(\textsection \ref{sec:eval-fault}) Saturation throughput for varying router fault rates.}}
\label{fig:routerfault-tp}
\vspace{-0.25em}
\end{figure*}

\Cref{fig:linkfault-tp,fig:routerfault-tp} show that LASH-TOR achieves the highest throughput for most fault scenarios, with (PA)ISF being a close second.
The throughput drop of (PA)ISF under faults indicates that missing path diversity significantly limits its performance.
Note that XY routing is not evaluated under faults, as they break the 2D mesh structure on which XY depends.

\subsubsection{Varying the Number of Virtual Channels}
\label{sec:eval-numvcs}

\begin{figure*}[h!]
\centering
\captionsetup{justification=centering}
\includegraphics[width=1.0\textwidth]{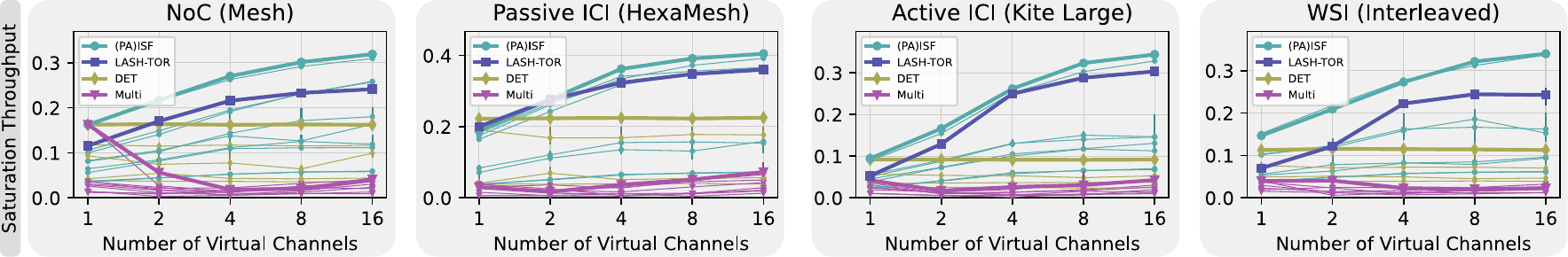}
\vspace{-1.5em}
\caption{{(\textsection \ref{sec:eval-numvcs}) Saturation throughput for varying \gls{vc} counts. A network with one \gls{vc} corresponds to a network that does not support \glspl{vc}.}}
\label{fig:numvcs-tp}
\vspace{-0.25em}
\end{figure*}

\Cref{fig:numvcs-tp} shows that (PA)ISF and LASH-TOR efficiently utilize \glspl{vc}, with throughput increasing significantly as more \glspl{vc} are added.
In contrast, \gls{det}'s performance remains unchanged with varying \glspl{vc}.
\textit{Multi}'s performance over varying \glspl{vc} exhibits a slight U-shape.

\subsubsection{Varying the Number of Entropy Bits in (PA)ISF}
\label{sec:eval-entropy}

\begin{figure*}[h!]
\centering
\captionsetup{justification=centering}
\includegraphics[width=1.0\textwidth]{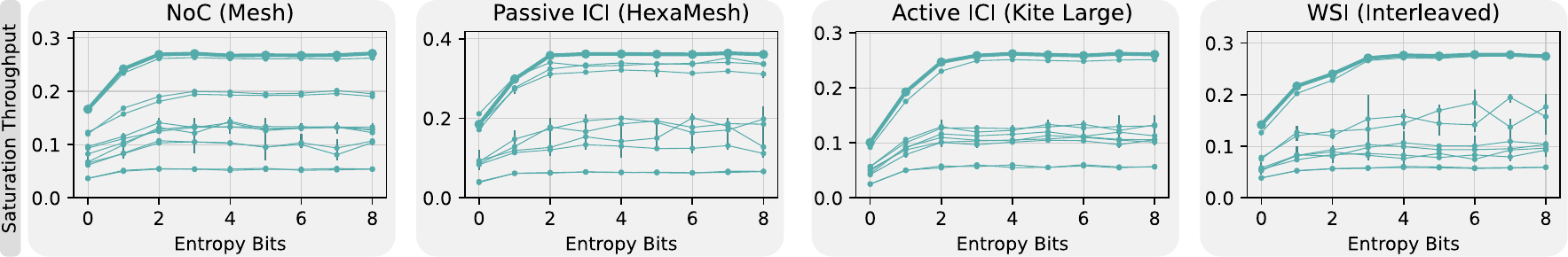}
\vspace{-1.5em}
\caption{{(\textsection \ref{sec:eval-entropy}) Saturation throughput for varying numbers of entropy bits per packet. Zero entropy bits results in deterministic selection.}}
\label{fig:entropy-tp} 
\end{figure*}

In \Cref{fig:entropy-tp}, increasing the number of entropy bits from zero (deterministic selection) to two significantly improves throughput, while using more than three bits yields almost no additional gain.
This is expected, as three bits encode eight values, and most topologies have fewer than eight neighbors per router to choose from.

\section{Conclusion}
\label{sec:conc}

\glsresetall

The increasing number of compute elements per chip, combined with packaging innovations such as chiplets and \gls{wsi}, has led to numerous specialized, irregular topologies for \glspl{noc}, \glspl{ici}, and \gls{wsi}.
Many applications require not only low-latency, high-throughput packet delivery, but also in-order delivery.
In this paper, we propose \textbf{ISF} and \textbf{PAISF}, two novel \glspl{sf} that guarantee in-order packet delivery.
We compare them against three existing methods for in-order delivery: a deterministic \gls{rf} (LASH-TOR), a deterministic \gls{sf} (DET), and the \textit{Multi} middle layer that sits between the \gls{rf} and \gls{sf}.
Our extensive evaluation across a range of topologies for \glspl{noc}, \glspl{ici}, and \gls{wsi}, using both synthetic and trace-based workloads, shows that ISF and PAISF achieve a 15.2\% increase in throughput, an 11.3\% reduction in latency under load, and a 12.1\% reduction in energy per transmitted packet, at the cost of a minor 3.3\% increase in zero-load latency.
These results demonstrate that ISF and PAISF provide an efficient and practical solution for high-performance, in-order packet delivery in modern irregular \gls{noc}, \gls{ici}, and \gls{wsi} topologies.

\ifnb
\section*{Acknowledgements}
\label{sec:ack}

This work was supported by the ETH Future Computing Laboratory (EFCL), financed by a donation from Huawei Technologies.
It also received funding from the European Research Council \raisebox{-0.25em}{\includegraphics[height=1em]{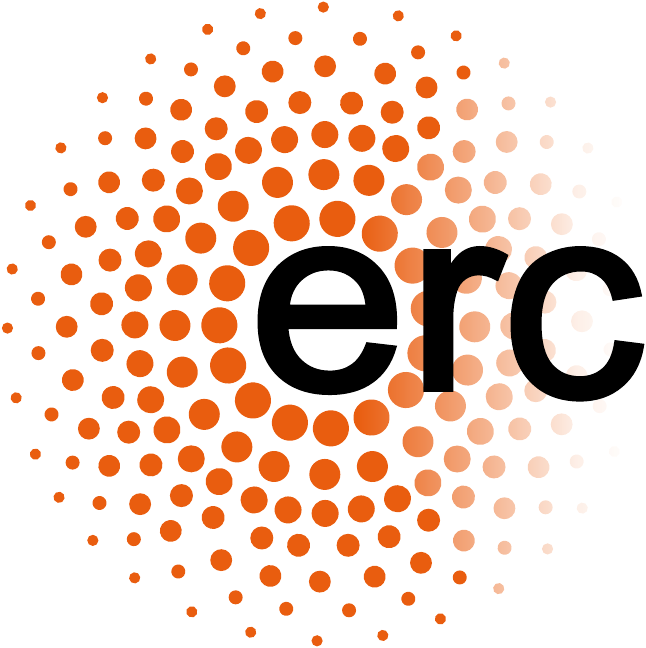}} (Project PSAP, No.~101002047).

\fi

\balance
\bibliographystyle{IEEEtran}
\bibliography{bibliography}

\end{document}